\documentclass{jpsj-suppl}
\usepackage{txfonts}

\title{Nucleon Structure Functions at Small $x$ via Holographic Pomeron Exchange}

\author{Akira \textsc{Watanabe}$^{1}$ and Katsuhiko \textsc{Suzuki}$^{2}$}

\inst{$^{1}$Institute of Physics, Academia Sinica, Taipei 11529, Taiwan, Republic of China \\
$^{2}$Department of Physics, Tokyo University of Science, Shinjuku, Tokyo 162-8601, Japan}

\email{watanabe@phys.sinica.edu.tw}

\recdate{September 15, 2015}

\abst{
The analysis on nucleon structure functions at small Bjorken $x$ in the framework of holographic QCD is presented.
In the model setup, the complicated nonperturbative interaction between the virtual photon and the target nucleon is described via the Pomeron exchange, which corresponds to the reggeized graviton exchange in the AdS space.
We show that our calculations for both $F_2$ and $F_L$ structure functions are in agreement with the experimental data measured at HERA.
}

\kword{deep inelastic scattering, holographic QCD, gauge/string correspondence}

\begin{document}
\maketitle

\section{Introduction}
A solid understanding of the quark-gluon structure of a nucleon has been one of the most important problems in high energy physics for several decades.
The structure functions, which can be measured via the lepton-nucleon deep inelastic scattering (DIS), are useful physical quantities to investigate it, and they are expressed by two kinematical variables, the Bjorken scaling variable $x$ and the photon four-momentum squared $Q^2$.
These two kinematical variables actualize the complicated dynamics of quarks and gluons in a nucleon.
If $x$ is not small ($x > 0.01$) and $Q^2$ is large enough ($Q^2 > 1$~GeV$^2$), perturbative QCD approaches are available, and assuming the initial conditions of parton distribution functions, one can predict the cross sections.
If $x$ is not small but $Q^2$ is small, we need to consider the hadron degree of freedom rather than that of quarks.
In the small $x$ region, since so many gluons with tiny momenta give a dominant contribution, studies based on effective models are basically needed, while the resummation technique has been developed in hard processes.
At small $x$, it is known that the Pomeron exchange, which is interpreted as a color singlet gluonic object, can well describe the total cross sections of two-body elastic scattering.

Focusing on the typically nonperturbative kinematical region, where $x < 0.01$ and $Q^2 <$ a few~GeV$^2$, here we present our analysis on the nucleon structure functions in the framework of holographic QCD based on our recent papers~\cite{Watanabe:2012uc,Watanabe:2013spa}.
The AdS/CFT correspondence (more generally, gauge/string correspondence) has gathered broad theoretical interests not only in high energy physics, but also in condensed matter physics, and various phenomenological applications have been attempted to study strongly coupled systems.
As for DIS, Polchinski and Strassler performed string calculations for the structure functions~\cite{Polchinski:2002jw}.
After that, many elaborated studies have been done~\cite{Brower:2006ea,Cornalba:2007zb,Brower:2007qh}, and especially Brower, Polchinski, Strassler, and Tan (BPST) proposed the Pomeron exchange kernel which gives the Pomeron exchange contribution to cross sections~\cite{Brower:2006ea}.
With the appropriate density distributions of the incident and target particles in the five-dimensional AdS space, the two-body scattering amplitude can be described with this kernel.
Several applications of the BPST kernel to DIS at small $x$ have been done so far, and the results in Refs.~\cite{Watanabe:2012uc,Watanabe:2013spa,Brower:2010wf,Watanabe:2015mia} are consistent with the experimental data.

To evaluate the structure functions, we need to specify the density distributions of the involved particles in the AdS space.
However, description of the nucleon is still under discussion, and various holographic models have been proposed in this decade.
We follow the analysis done by Abidin and Carlson~\cite{Abidin:2009hr}, and calculate the density distribution, which is identified as the gravitational form factor in their study.
In this proceedings, we show in the next section the nucleon density distributions with the hard- and soft-wall AdS/QCD models, and discuss the difference between them.
After that, we show that our calculations for the structure functions with the soft-wall model, where the AdS geometry is smoothly cut off at the infrared (IR) region, are in agreement with the HERA data (see Ref.~\cite{Watanabe:2013spa} for details).
Also, we remark on the feasibility of more applications with this framework to other high energy scattering processes, in which the Pomeron exchange gives dominant contribution.

\section{Model setup}
Following the previous studies~\cite{Watanabe:2012uc,Watanabe:2013spa,Brower:2010wf} and considering the single Pomeron exchange, the structure functions are expressed with the BPST kernel $\chi$ as
\begin{equation}
F_i (x,Q^2) = \frac{g_0^2 \rho^{3/2}Q^2}{32 \pi ^{5/2}} \int dzdz' P_{13}^{(i)} (z,Q^2) P_{24}(z') (zz') \mbox{Im} [\chi (s,z,z')], \label{eq:Fi}
\end{equation}
where $i = 2, L$ are for $F_2$ and $F_L$ structure functions, respectively, $s$ is the usual Mandelstam variable, i.e., $x \approx Q^2/s$, and $z(z')$ are the bulk coordinates.
$g_0$ is an overall model parameter which governs the magnitude of the structure functions.
Another adjustable parameter $\rho$ appears in the Pomeron exchange kernel also, and controls the energy dependence.
$P_{13(24)}$ are density distributions of the involved particles for the two-body scattering, $1 + 2 \to 3 + 4$, in the AdS space, and they affect the property of the Pomeron.
The exchanged Pomeron becomes softer with the overlap of the two density distributions increasing, e.g., hadron-hadron elastic scattering~\cite{Donnachie:1992ny}, and it becomes harder with the overlap decreasing, e.g., DIS with $Q^2 > 1$~GeV$^2$~\cite{Breitweg:1998dz}.

As long as we concentrate on the single Pomeron exchange, the kernel part of Eq.~\eqref{eq:Fi} can be expressed in a closed form~\cite{Brower:2007qh}.
Originally, two kinds of kernel were proposed.
One of them is the conformal kernel which was rigorously derived in the conformal limit, and the other is the modified kernel, in which the authors of Ref.~\cite{Brower:2006ea} tried to mimic the nonperturbative confinement effect of QCD.
In the following sections, we only discuss the results with the modified kernel.

\begin{figure}[tb]
\begin{center}
\includegraphics[width=0.99\textwidth]{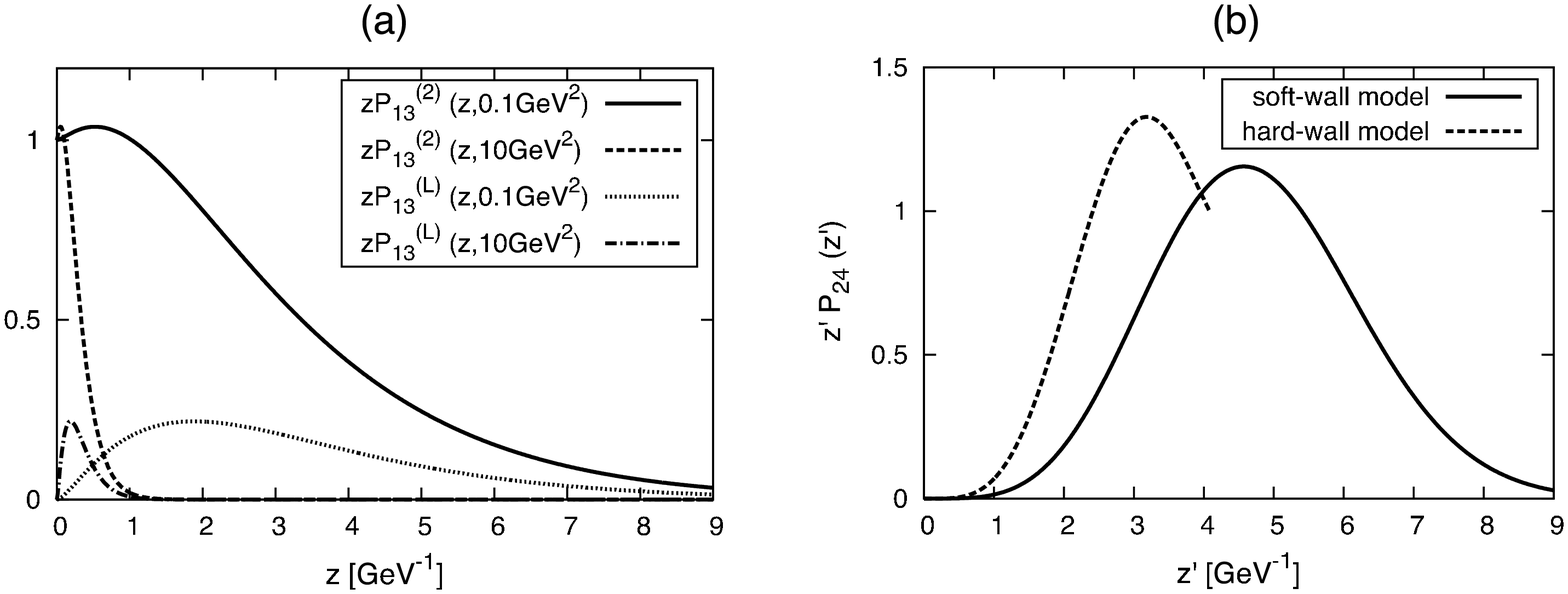}
\caption{
Density distributions of the involved particles in the $z(z')$ space.
(a) $zP_{13}^{(2)}(z,Q^2)$ with $Q^2 = 0.1, 10$~GeV$^2$ shown by solid and dashed curves, and $zP_{13}^{(L)}(z,Q^2)$ with $Q^2 = 0.1, 10$~GeV$^2$ indicated by dotted and dash-dotted curves, respectively.
(b) $z'P_{24}(z')$ with the soft- and hard-wall models depicted by solid and dashed curves, respectively.
}
\label{OF}
\end{center}
\end{figure}

To evaluate the structure functions by Eq.~\eqref{eq:Fi}, we need to specify the density distributions $P_{13}$ and $P_{24}$.
For the density distribution of the probe photon, we utilize the wave function of the 5D U(1) vector field in the AdS space, which is identified as the physical photon at the ultraviolet (UV) boundary~\cite{Polchinski:2002jw}.
$P_{13}^{(2)}$ includes both the transverse and longitudinal components which is used to calculate the $F_2$ structure function, and $P_{13}^{(L)}$ only includes the longitudinal component which is used to evaluate the longitudinal structure function $F_L$.
Since the integrand of Eq.~\eqref{eq:Fi} includes $zz'$, we plot $zP_{13}^{(2)}$ and $zP_{13}^{(L)}$ in Fig.~\ref{OF}~(a)
to see their behavior in the $z$ space.
$zP_{13}^{(2)}$ is localized around the origin when $Q^2$ is large ($> 1$~GeV$^2$), and it spreads broadly from the UV to IR region when $Q^2$ is small ($< 1$~GeV$^2$).
The behavior of $zP_{13}^{(L)}$ is similar to that of $zP_{13}^{(2)}$.

The density distribution of the target nucleon is extracted from the nucleon-Pomeron-nucleon three point function, which can be realized by utilizing holographic models of the nucleon.
Following the preceding study~\cite{Abidin:2009hr}, one can obtain the gravitational form factor from the 5D classical action of the AdS/QCD model proposed by Hong, Inami, and Yee~\cite{Hong:2006ta}.
In the model, the nucleon is expressed as a 5D Dirac fermion with chiral symmetry breaking.
One can consider two kinds of model action, the hard- and soft-wall models, and here we need to consider either one to obtain the density distribution.
In the previous papers, we tried to consider both models, and found that the density distributions from them are obviously different, which are displayed in Fig.~\ref{OF}~(b).
From the figure, one can see the long tail of the distribution obtained with the soft-wall model in the IR region, while the hard-wall one is cut off at $z' \sim 4$~GeV$^{-1}$, which corresponds to the typical size of the nucleon, $0.8$~fm.

\section{Numerical results and discussion}
\begin{figure}[tb]
\begin{center}
\includegraphics[width=0.99\textwidth]{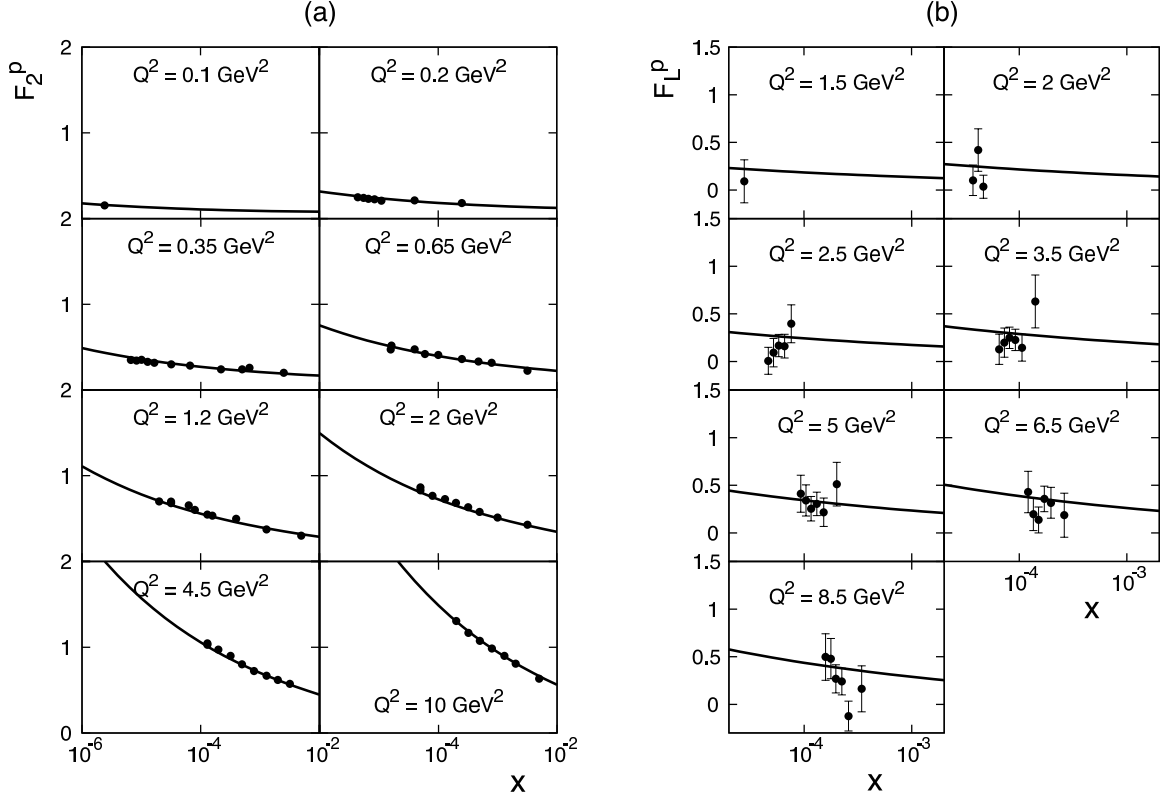}
\caption{
(a) $F_2^p (x,Q^2)$ and (b) $F_L^p (x,Q^2)$ as a function of the Bjorken $x$ for various $Q^2$.
In each panel, the solid curves represent our calculations with the modified kernel and the nucleon density distribution obtained by the soft-wall AdS/QCD model.
HERA data for $F_2$~\cite{Aaron:2009aa} and $F_L$~\cite{Collaboration:2010ry} structure functions are depicted by circles with error bars.
}
\label{F2_and_FL}
\end{center}
\end{figure}

Combining the two density distributions and the BPST Pomeron exchange kernel, and fixing three adjustable parameters, $g_0$, $\rho$, and $z_0$ which appears in the modified kernel and controls the strength of the confinement effect, one can evaluate the structure functions.
We show in Fig.~\ref{F2_and_FL}~(a)
our calculations with the modified kernel and the nucleon density distribution obtained by the soft-wall model for $F_2$ structure function.
Our results are quite consistent with the experimental data measured at HERA in the considered kinematical region, which mean that our model setup describes both the $x$ and $Q^2$ dependence correctly.

Once the three adjustable parameters are fixed in the analysis on the $F_2$ structure function, we can also evaluate the longitudinal structure function without any additional parameter.
We compare our predictions of $F_L$ with HERA data, which is shown in Fig.~\ref{F2_and_FL}~(b).
Although the $F_L$ data have somewhat large errors, one can see that our results are in agreement with the data within the error bars.
Furthermore, we considered the so-called longitudinal-to-transverse ratio, which is defined as $R = F_L(x,Q^2) / F_T(x,Q^2)$ where $F_T = F_2 - F_L$ (see Fig.~7 in Ref.~\cite{Watanabe:2013spa}).
We found that the ratio $R$ increases with the Bjorken $x$ and/or the scale $Q^2$.
The $Q^2$ dependence is substantial, while the $x$ dependence is comparatively weak.
It should be noted that the qualitative behavior of our results is consistent with that of the DGLAP fit in the ACOT scheme (see Fig.~14 in Ref.~\cite{Collaboration:2010ry}).
Behavior of the ratio $R$ is extremely nontrivial in the small-$x$ region, and we need more precise experimental data for $F_L$ to draw a conclusion.

In conclusion, we have studied the nucleon structure functions in the framework of holographic QCD, focusing on the small Bjorken $x$ region.
The nontrivial $x$ and $Q^2$ dependence of the experimental data for the $F_2$ structure function are well described in our model, and our predictions of the longitudinal structure function are in agreement with the data measured at HERA.
If the Pomeron exchange gives dominant contribution in certain kinematical region, once the appropriate density distributions of the involved particles are given, any two-body scattering process at high energy can be considered in this framework.
Besides the nucleon, the pion structure functions~\cite{Watanabe:2012uc} and the photon structure functions~\cite{Watanabe:2015mia} have been studied so far, and their calculations are consistent with the experimental data.
All these results may show the universality of the Pomeron in high energy scattering.
Further investigations are certainly needed.



\end{document}